\documentclass[twocolumn,showpacs,aps,preprintnumbers,amsmath,amssymb]{revtex4}

\usepackage{graphicx,subfigure,multirow}
\usepackage{dcolumn}
\usepackage{bm}
\usepackage{axodraw4j}
\usepackage[all]{xy}

\makeatletter

\newcommand{\fmslash}[2][0mu]{%
  \mathchoice
    {\fmsl@sh\displaystyle{#1}{#2}}%
    {\fmsl@sh\textstyle{#1}{#2}}%
    {\fmsl@sh\scriptstyle{#1}{#2}}%
    {\fmsl@sh\scriptscriptstyle{#1}{#2}}}
\newcommand{\fmsl@sh}[3]{%
  \m@th\ooalign{$\hfil#1\mkern#2/\hfil$\crcr$#1#3$}}
\makeatother

\newcommand{\lsim}{{\;\raise0.3ex\hbox{$<$\kern-0.75em\raise-1.1ex\hbox{$\sim$}}\;}}
\newcommand{\gsim}{{\;\raise0.3ex\hbox{$>$\kern-0.75em\raise-1.1ex\hbox{$\sim$}}\;}}

\newcommand{\met}{{\fmslash E_T}}
\usepackage{color}

\begin{document}
\title{How to look for supersymmetry under the lamppost at the LHC}

\author{Partha Konar${}$, Konstantin T.~Matchev${}$, Myeonghun Park${}$, and Gaurab K.~Sarangi${}$}
\affiliation{${}$Physics Department, University of Florida, Gainesville, FL 32611, USA}

\date{August 14, 2010}

\begin{abstract}
We apply a model-independent, agnostic approach to the 
collider phenomenology of supersymmetry (SUSY),
in which {\em all} mass parameters are taken 
as free inputs at the weak scale. We consider the gauginos,
higgsinos, and the first two generations of sleptons and squarks, 
and analyze all possible mass hierarchies among them
($4\times 8!=161,280$ in total) in which the 
lightest superpartner
is neutral, leading to missing energy.
In each case, we identify the full set of 
the dominant (i.e.~least suppressed by phase space, small 
mixing angles or Yukawas) decay chains originating from the 
lightest colored superpartner.
Our exhaustive search reveals several quite dramatic yet 
unexplored multilepton signatures with up to 8 isolated leptons
(plus possibly up to 2 massive gauge or Higgs bosons)
in the final state. Such events are spectacular, 
background-free for all practical purposes, and may
lead to a discovery in the very early stage ($\sim 10\ {\rm pb}^{-1}$) of LHC operations at 7\,TeV.
\end{abstract}





\maketitle

The ramping operations at the CERN Large Hadron Collider (LHC) 
have begun the long awaited and historic exploration of the TeV 
scale, where new physics beyond the standard model (SM) 
is expected to emerge. Among the multitude of new scenarios, low energy 
supersymmetry (SUSY) has long been the primary target of the LHC
experiments, not just because it is well motivated
theoretically \cite{Chung:2003fi}, but also because its
generic discovery signatures cover a much wider class of models
\cite{Cheng:2002ab}.

By itself, SUSY is very predictive, as it fixes the spins and
couplings of the new particles (the superpartners). 
Unfortunately, this is not sufficient to pin down its 
precise collider discovery signatures, as the latter
crucially depend on the SUSY mass spectrum, which is in turn
determined by the mechanism of supersymmetry {\em breaking}.
Alas, almost 40 years of model building effort since the discovery
of supersymmetry have failed to produce a single, universally 
accepted model of SUSY breaking. 

Given one's utter ignorance about the expected pattern of SUSY 
masses, in this letter we adopt a most conservative,
agnostic approach, where the masses of all superpartners are 
treated as free inputs at the weak scale. We shall then 
consider {\em all} possible hierarchical patterns among them,
and identify the set of dominant (in the sense defined below) 
collider signatures in each case.
In our quest for interesting models, we shall be guided by
experimental pragmatism instead of theoretical prejudice.

The purpose of this study is twofold. First, most previous 
collider studies of SUSY have been performed for specific SUSY 
benchmark points, typically chosen within some minimal model 
such as ``minimal supergravity'' (MSUGRA) \cite{Battaglia:2001zp}. 
We will therefore be interested in uncovering new types of signatures 
which may have been missed in the standard benchmark approach.
Secondly, we shall focus our search on
signatures with a high number of isolated leptons, which
constitute the proverbial ``smoking gun'' for new physics.
For example, the inclusive trilepton channel is already recognized
as ``the golden mode'' for an early SUSY discovery at hadron colliders.
One of our main results here will be the identification of 
a number of new SUSY mass patterns whose {\em dominant} signatures
have up to {\em eight} leptons in the final state.

\begin{table}[t]
\caption{\label{tab:summary}
The set of SUSY particles considered in this analysis, shorthand notation
for each multiplet, and the corresponding soft SUSY breaking mass parameter.
}
\begin{ruledtabular}
\begin{tabular}{| c | c | c | c | c | c | c | c | c | c |}
 $\tilde{u}_L$,  $\tilde{d}_L$ & $\tilde{u}_R$ & $\tilde{d}_R$ & $\tilde{e}_L$,  $\tilde{\nu}_L$ & $\tilde{e}_R$ & $\tilde{h}^\pm$,  $\tilde{h}_u^0$,  $\tilde{h}_d^0$ & $\tilde{b}^0$ & $\tilde{w}^\pm$,$\tilde{w}^0$ & $\tilde{g}$ \\ \hline\hline
$Q$ & $U$ & $D$ & $L$ & $E$ & $H$ & $B$ & $W$ & $G$ \\ \hline
$M_Q$ & $M_U$ & $M_D$ & $M_L$ & $M_E$ & $M_H$ & $M_B$ & $M_W$ & $M_G$ 
\end{tabular}
\end{ruledtabular}
\end{table}

Our setup is as follows. We take the usual superpartner content 
of the Minimal Supersymmetric Standard Model (MSSM) 
listed in Table~\ref{tab:summary}. For simplicity, 
in this letter we shall consider just two degenerate
light generations of sfermions. Third generation effects
can be trivially incorporated in the discussion \cite{KMPS}, 
and only complicate the bookkeeping.
Given the 9 input mass parameters in Table~\ref{tab:summary}, 
in general there are $9!=362,880$ possible
orderings among them, each leading to a 
distinct pattern (hierarchy) of sparticle masses.
We shall use the shorthand notation from Table~\ref{tab:summary}
to label each hierarchy: for example, $GQUDHLWEB$ is a model with 
$M_G > M_Q > M_U > M_D > M_H > M_L > M_W > M_E > M_B$.

Our first goal will be to identify the main
collider signatures for each hierarchy.
As in any discussion on SUSY collider phenomenology,
our starting point is the fate and then the nature 
of the Lightest Supersymmetric Particle 
(LSP), which we shall generically denote by ${\cal L}$.
For our main analysis, we shall assume that $R$-parity is conserved 
(or very weakly broken), so that ${\cal L}$ is stable on the
scale of the detector. (We briefly discuss the $R$-parity 
violating option at the end.) 
Then, the original $9!$ model hierarchies can be classified into 
the following three categories: 

I. {\em CHAMPs.} In the $8!=40,320$ cases with ${\cal L}=E$,
the LSP is an electrically charged, color-neutral particle
(the right-handed slepton $\tilde e_R$). The corresponding
generic collider signature is a long-lived charged massive particle 
(CHAMP) \cite{CHAMP}, regardless of the particular ordering of the
heavier sparticles. 

II. {\em R-hadrons.} In $4\times 8!=161,280$ of the remaining
hierarchies ${\cal L}\in \{Q,U,D,G\}$, the LSP is a colored particle,
and the generic searches for stable R-hadrons apply \cite{Kraan:2005ji}.
Again the ordering of the heavier particles is not
particularly important.

III. {\em Missing transverse energy.} In the remaining 
$4\times 8!=161,280$ cases ${\cal L}\in \{L,B,W,H\}$ and
the LSP is a weakly-interacting, electrically 
neutral particle. Its production 
will lead to missing transverse energy ($\met$) in the 
detector. Now, however, the signatures crucially depend 
on the ordering of the heavier particles, since 
it is not feasible to look for $\met$ inclusively.
Our goal here is to fully classify these
$161,280$ models according to their collider 
phenomenology. Unlike previous general approaches
\cite{Feldman:2007zn}, which employed scans of the 
multi-dimensional SUSY parameter space, here 
we would like to avoid scanning, keeping the discussion simple and 
qualitative.

\begin{figure}[t]
\includegraphics[width=6.5cm]{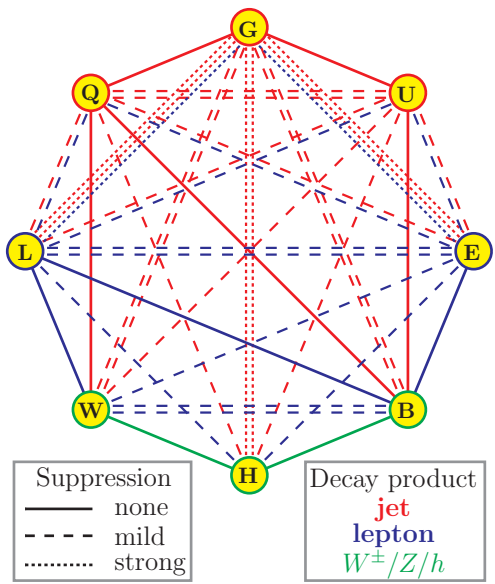} 
\caption{\label{fig:flowchart}
Graphical representation of the allowed transitions between 
the SUSY states from Table~\ref{tab:summary}. 
One (two, three) parallel lines represent 
two- (three-, four-) body decays.
The relative suppression of each decay mode is 
indicated by the line type. The identity
of the resulting SM decay products is denoted by the
line color: red for a jet $j$, blue for a lepton $\ell$
and green for a massive boson $v\equiv\{W^\pm,Z,h\}$ 
(which may be either on-shell or off-shell). }
\end{figure}

Both of the currently operating high energy colliders 
(the Tevatron at Fermilab and the LHC at CERN)
are {\em hadron} machines, at which the total production
is expected to be dominated by the strong production
of colored superpartners. Correspondingly, the starting point
for our classification will be the nature 
of the lightest colored superpartner (LCP), denoted by ${\cal C}$.
Then, each of the $161,280$ missing energy hierarchies 
at hand can be represented by a particular ordering 
\begin{equation}
x\dots x\,{\cal C}\,y\dots y\,{\cal L}\, ,
\label{hi}
\end{equation}
where the $x$'s stand for inconsequential entries,
${\cal C}\in \{G,Q,U,D\}$, $y\in \{L,B,W,H,E\}$
and ${\cal L}\in \{L,B,W,H\}$.
The dominant collider signature for each model hierarchy (\ref{hi}) 
will be determined by the {\em inclusive} pair production of 
${\cal C}$ and its {\em dominant} subsequent decays. 

Our key idea here is that once a given hierarchy (\ref{hi}) is assumed,
the dominant decay modes of ${\cal C}$ are uniquely determined,
since supersymmetry predicts all superpartner couplings.
In our analysis, we shall assume that there are no accidental 
phase space suppressions due to any two mass parameters 
from Table~\ref{tab:summary} being very close. This assumption 
also guarantees that the chargino and neutralino mixing angles are 
small and the mass eigenstates are roughly aligned with the
interaction eigenstates. One can then use the simple 
chart in Fig.~\ref{fig:flowchart} to identify the
dominant (i.e.~least suppressed) decay modes of ${\cal C}$, 
which we label by the number of leptons $n_\ell$ (blue lines),
number of jets $n_j$ (red lines) and number of massive bosons $n_v$ 
(green lines) encountered along the way. 
Solid lines in the figure correspond 
to 2-body decays which do not suffer from any 
(chargino or neutralino) mixing angle suppression (MAS);
dashed lines indicate either 2-body decays with MAS
or 3-body decays with no MAS; and finally, dotted lines 
stand for either 3-body decays with MAS
or 4-body decays with no MAS. We then count
the number of mass hierarchies (\ref{hi}) 
which exhibit a dominant decay channel for ${\cal C}$ 
with a given set of $(n_\ell,n_v,n_j)$, and 
show the result in Table~\ref{tab:signature}.
\begin{table}[t]
\caption{\label{tab:signature} 
Number of hierarchies for the various dominant decay modes
of the LCP ${\cal C}$.}
\begin{ruledtabular}
\begin{tabular}{| c | c | c | c | c | c | c |}
     & \multicolumn{2}{c|}{$n_v=0$} & \multicolumn{2}{c|}{$n_v=1$} & \multicolumn{2}{c|}{$n_v=2$} \\ \hline
$n_\ell$ & $n_j=1$ & $n_j=2$ & $n_j=1$ & $n_j=2$ & $n_j=1$ & $n_j=2$ \\ 
\hline
 $0$ & 79296 & 26880 & 12768 & 3360 & 1344 & 672 \\
 $1$ & 30240 & 10080 &  1824 &  480 &  192 &  96 \\
 $2$ & 19770 &  6030 &  1500 &  180 &    0 &   0 \\ 
 $3$ &  4656 &  1296 &   312 &   72 &    6 &   6 \\
 $4$ &  1656 &   396 &    66 &    6 &    0 &   0 \\
\end{tabular}
\end{ruledtabular}
\end{table}
At times, there can be several dominant decay modes of ${\cal C}$.
For example, consider $xxxxQWBLH$. One can get 
$(n_\ell,n_v,n_j)=(0,1,1)$ in two ways:
$Q\to W\to H$ or $Q\to B\to H$. It is also possible to have
$(n_\ell,n_v,n_j)=(2,0,1)$ in two different ways:
$Q\to W\to L\to H$ or $Q\to B\to L\to H$.
Therefore, $xxxxQWBLH$ contributes one entry to 
each of the two boxes $(n_\ell,n_v,n_j)=(0,1,1)$
and $(n_\ell,n_v,n_j)=(2,0,1)$ in Table~\ref{tab:signature}.
As a result, the total number of entries (203,184)
in Table~\ref{tab:signature}
is larger than the total number of hierarchies (161,280).

Table~\ref{tab:signature} leads to some interesting 
conclusions. For example, we see that the purely hadronic 
signatures of $n_j$ jets and $\met$ alone cover a very large
fraction ($\sim 65\%$) of all possible SUSY hierarchies
with a neutral LSP. There is also a sizable fraction of 
models which can be explored via the standard searches
for signatures with one, two or three leptons. 
Keep in mind that the LCP's are produced 
in pairs, so the collider signature is obtained by
{\em doubling} the number of leptons and jets displayed 
in the table and is given by $2n_\ell+2n_v+2n_j+\met$.
Perhaps the most important result from 
Table~\ref{tab:signature} is that there is a non-negligible
fraction of SUSY models ($\sim 1\%$) in which one of
the dominant decay modes of the LCP yields 4 isolated 
leptons, and the corresponding collider signature is
8 leptons plus $\met$! 

\begin{table}
\caption{\label{tab:signatureL} 
Number of hierarchies for the {\em maximally leptonic}
decay modes of the LCP ${\cal C}$.}
\begin{ruledtabular}
\begin{tabular}{| c | c | c | c | c | c | c |}
     & \multicolumn{2}{c|}{$n_v=0$} & \multicolumn{2}{c|}{$n_v=1$} & \multicolumn{2}{c|}{$n_v=2$} \\ \hline
$n_\ell$ & $n_j=1$ & $n_j=2$ & $n_j=1$ & $n_j=2$ & $n_j=1$ & $n_j=2$ \\ 
\hline
 $0$ & 61488 & 21168  &  8310 & 2550 &  780 & 420 \\
 $1$ & 24150 &  8310  &  1278 &  378 &  132 &  72 \\
 $2$ & 17190 &  5550  &  1230 &  150 &    0 &   0 \\ 
 $3$ &  4362 &  1242  &   312 &   72 &    6 &   6 \\
 $4$ &  1656 &   396  &    66 &    6 &    0 &   0 \\
\end{tabular}
\end{ruledtabular}
\end{table}

We now repeat the same analysis, only this time 
from the set of all dominant decay modes of ${\cal C}$,
we select the one with the largest $n_\ell$, 
and in case of a tie for $n_\ell$, we pick the 
chain with the larger $n_v$. We refer to 
such decay modes of the LCP as ``maximally leptonic''.
The new tally is displayed in Table~\ref{tab:signatureL},
where now the sum of all entries equals the total number 
of signatures 161,280.

\begin{figure}[b]
\includegraphics[width=7.0cm]{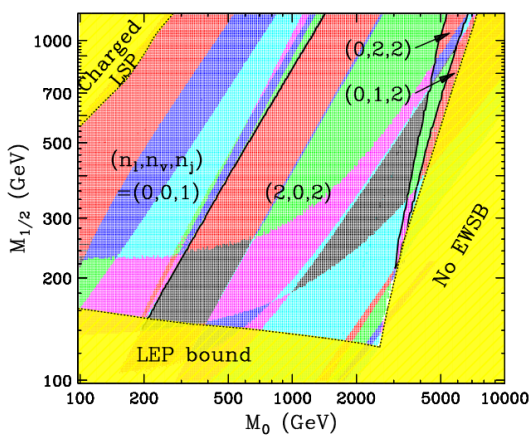} 
\caption{\label{fig:sugra}
A slice through the MSUGRA parameter space for fixed
$A_0=0$, $\tan\beta=10$ and $\mu>0$. Yellow shaded regions 
are ruled out by direct searches at LEP
or by requiring a neutral LSP. The remaining area
is color-coded according to the sparticle hierarchy type (\ref{hi}).
The solid black lines delineate regions with the same maximally
leptonic ${\cal C}$ decay mode.}
\end{figure}

It is instructive to apply our general formalism 
to the familiar MSUGRA model, where
the soft SUSY mass parameters of Table~\ref{tab:summary}
have common values ($M_0$ for the scalar superpartners and 
$M_{1/2}$ for the gauginos) at the grand unification scale. 
First, one may ask how 
many of the 161,280 missing energy hierarchies 
are actually represented in MSUGRA. The answer is provided in
Fig.~\ref{fig:sugra}, in which we divide the 
usual $(M_0,M_{1/2})$ plane into disjoint color-coded areas,
according to the observed mass pattern (\ref{hi}).
The allowed region contains 
only 47 different hierarchies (some areas are too 
small to be readily noticeable with the naked eye).
Comparing this to the total number of 
161,280 possibilities, one gets an idea of the 
limitations of MSUGRA as a benchmark scenario.

We also use Fig.~\ref{fig:sugra} to illustrate the extent to which 
the MSUGRA model is able to cover the generic SUSY signature space. 
The black solid lines in Fig.~\ref{fig:sugra}
divide the allowed portion of the  
$(M_0,M_{1/2})$ plane into disjoint regions, classified according 
to the maximally leptonic ${\cal C}$ decay mode, labelled by $(n_\ell,n_v,n_j)$.
We find that MSUGRA exhibits only 4 out of the 26 
possibilities found in Table~\ref{tab:signatureL}.
At small values of $M_0$, the LCP is a 
right-handed squark (either $\tilde u_R$ or $\tilde d_R$),
whose single dominant decay mode is directly to the 
bino-like LSP: 
$\{\tilde u_R, \tilde d_R \} \stackrel{j}{\to} \tilde b^0$. 
As the value of $M_0$ increases and the squarks get heavier, 
the gluino eventually becomes the LCP and its maximally 
leptonic decay mode is 
$\tilde g \stackrel{jj}{\to} \tilde w^0 \stackrel{\ell\ell}{\to}\tilde b^0$.
Upon further increasing $M_0$, we eventually
enter the focus point region \cite{Feng:1999zg}, where 
at first $M_W > M_H > M_B$ and the winos decay as
$\{\tilde w^{\pm},\tilde w^0\} \stackrel{v}{\to} 
\{\tilde{h}^\pm,\tilde{h}_u^0,\tilde{h}_d^0\}
\stackrel{v}{\to}\tilde b^0$, while in the rightmost 
portion of the plot one finds
$M_W > M_B > M_H$ with the direct transitions 
$\{\tilde w^{\pm},\tilde w^0\} \stackrel{v}{\to} 
\{\tilde{h}^\pm,\tilde{h}_u^0,\tilde{h}_d^0\}$.
Of course, Fig.~\ref{fig:sugra} only displays
{\em dominant} LCP decay modes. Some parameter 
points in MSUGRA may contain other, longer 
decay chains with a higher number of leptons, 
but those would be subdominant and therefore 
suppressed by branching fractions. For example,
the hierarchy $xxxUHLWEB$ has a 4-lepton chain
$\tilde u_R \stackrel{j}{\to} \tilde h^0
\stackrel{\ell}{\to} \tilde e_L
\stackrel{\ell}{\to} \tilde w^0
\stackrel{\ell}{\to} \tilde e_R
\stackrel{\ell}{\to} \tilde b^0$,
but it has two bottlenecks at 
$\tilde u_R \stackrel{j}{\to} \tilde h^0$ and
$\tilde h^0\stackrel{\ell}{\to} \tilde e_L$, see Fig.~\ref{fig:flowchart}.

\begin{table}[t]
\caption{\label{tab:mass}
Input soft SUSY mass parameters (in GeV) for the 
$xxGQWLBEH$ study points used for Figs.~\ref{fig:lepton_xsec}
and \ref{fig:leptoncount}.}
\begin{tabular}{| c | c | c | c | c | c | c | }
\hline
$~M_G~$&$~M_Q~$&$~M_W~$&$~M_L~$&$~M_B~$&$~M_E~$ &$~M_H~$ \\ \hline
 400 & 300 & 220 & 190 & 130 & 130 & 130 \\
 450 & 350 & 280 & 190 & 120 & 120 & 120 \\ 
 500 & 400 & 280 & 190 & 120 & 120 & 120 \\ 
 550 & 450 & 310 & 200 & 120 & 120 & 120 \\ 
 600 & 500 & 350 & 210 & 130 & 120 & 120 \\ 
 700 & 600 & 420 & 230 & 150 & 130 & 120 \\ 
 800 & 700 & 480 & 250 & 160 & 130 & 120 \\ 
 900 & 800 & 500 & 250 & 170 & 130 & 120 \\ 
1000 & 900 & 510 & 250 & 170 & 130 & 120 \\
\hline
\end{tabular}
\end{table}

\begin{figure}[b]
\includegraphics[width=6.5cm]{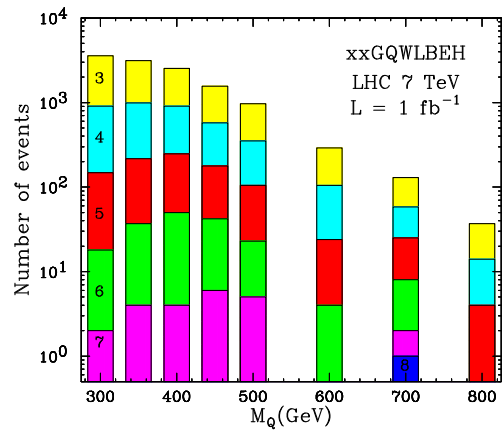} 
\caption{\label{fig:leptoncount}
(Stacked) number of multi-lepton events in 1 ${\rm fb}^{-1}$ of data at a 7 TeV
LHC, for the $xxGQWLBEH$ sparticle hierarchy with the mass spectra
shown in Table~\ref{tab:mass}.}
\end{figure}

In the rest of this letter we study in detail one
example\footnote{The complete list of all 161,280 hierarchies 
analyzed in this paper and their signatures
will be given in \cite{KMPS}.} 
of a maximally leptonic decay chain with 4 leptons, 
where the corresponding collider signal
is 8 isolated leptons plus jets plus missing energy.
For concreteness, consider the hierarchy $xxGQWLBEH$
at several different study points, defined in Table~\ref{tab:mass}
and chosen ``under the lamppost'', i.e.~to maximize 
the 8-lepton signal rate.
For this hierarchy, 8 lepton events arise from the inclusive
pair production of left-handed squarks $\tilde u_L, \tilde d_L$,
followed by
\begin{equation}
\tilde u_L, \tilde d_L \stackrel{j}{\to} 
\tilde w^0 \stackrel{\ell}{\to} 
\tilde \ell_L \stackrel{\ell}{\to} 
\tilde b^0 \stackrel{\ell}{\to} 
\tilde \ell_R \stackrel{\ell}{\to} 
\tilde h^0_u, \tilde h^0_d\ .
\label{8lep}
\end{equation}
Ignoring phase space suppression factors, 
third generation effects and the 
masses of the $W^\pm$, $Z$ and $h$, it is easy to 
estimate the individual branching ratios for this chain
as follows \cite{KMPS}: $BR(\tilde g\to \tilde q_L + j) = 1$;
$BR(\tilde q_L\to \tilde w^0 + j )\simeq \frac{1}{3}$;
$BR(\tilde w^0 \to \tilde \ell_L^\pm +\ell^\mp)\simeq \frac{1}{3}$;
$BR(\tilde \ell^\pm_L \to \tilde b^0 + \ell^\pm) = 1$;
$BR(\tilde b^0 \to \tilde \ell_R^\pm +\ell^\mp) \simeq \frac{4}{5}$;
$BR(\tilde \ell_R^\pm\to \tilde h^0_{u/d} + \ell^\pm)= 1$.
\begin{figure}[t!]
\includegraphics[width=7.0cm]{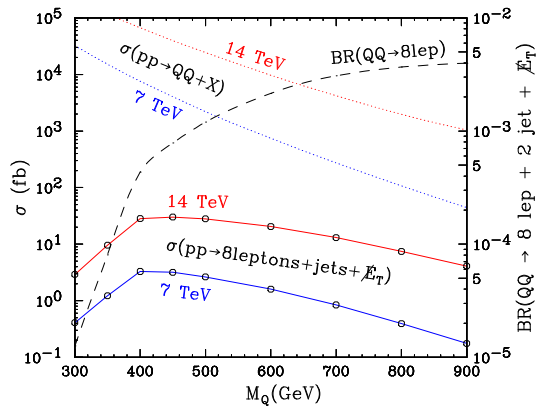} 
\caption{\label{fig:lepton_xsec}
Inclusive cross-sections (in fb) at the LHC for 8 lepton events 
from the exclusive chain in (\ref{8lep})
for the study points from Table~\ref{tab:mass} (solid lines);
inclusive cross-sections for left-handed 
squark pair production (dotted); and
cumulative branching fraction of the 8 lepton final state
(dashed, read on the right).}
\end{figure}
Multiplying these results 
and squaring, we expect the cumulative branching fraction for 
8 lepton events to be around 0.7\%. 
This is confirmed by the dashed line in Fig.~\ref{fig:lepton_xsec},
which approaches this asymptotic value for large mass splittings 
(large $M_Q$). We see that to maximize the overall branching fraction, 
we must 
scale the SUSY mass spectrum up. Yet to maximize the 
production rate, we must scale the spectrum down. 
An optimum compromise is therefore found for 
intermediate values of the SUSY masses, as shown in 
Fig.~\ref{fig:lepton_xsec}. The study points 
in Table~\ref{tab:mass} were picked by varying $M_Q$, 
fixing $M_G=M_Q+100$ GeV, and choosing the rest of the
spectrum from a coarse scan to maximize the 8 lepton 
rate shown by the solid lines in Fig.~\ref{fig:lepton_xsec}.
These study points were then processed with the PYTHIA
\cite{Sjostrand:2006za} event generator and
the PGS \cite{PGS} detector simulator for the case of 
an LHC at 7 TeV center-of-mass energy and just 
$1\ {\rm fb}^{-1}$ of data. Because these events are
essentially background free, we simply count
PGS-reconstructed isolated leptons with default $p_T$ cuts
of $3$ GeV for muons and $10$ GeV for electrons,
and display the result in Fig.~\ref{fig:leptoncount}.
Due to the imperfect detector acceptance, we got only one 
8 lepton event in $1\ {\rm fb}^{-1}$. Nevertheless,
for $M_Q \lsim 500$ GeV typically there are a handful 
of 7 lepton events and hundreds of 4 lepton events, which are already
very clean. Since only 5 clean events 
are sufficient for discovery, the region $M_Q \lsim 500$ GeV
can be probed with as little as $10\ {\rm pb}^{-1}$ of data.

We note in passing that one may expect even more leptons
if $R$-parity is violated and the LSP decays promptly
to SM particles. For example, lepton number violation 
of the type $LLE$ would lead to 2+2=4 additional leptons, 
bringing the maximum lepton total per event to 12.

In conclusion, we have shown that already with its 
first $10\ {\rm pb}^{-1}$ of data the LHC can start 
probing SUSY, provided one looks in the right place. 
To this end, 
it is imperative to search for the model hierarchies
that are {\em most likely} to be discovered first.
We believe this letter outlines the most general strategy to do just that.

\vspace{-0.5cm}
\acknowledgments
\noindent
Work supported by US DoE
grant DE-FG02-97ER41029.



\end{document}